\documentclass{article}

\usepackage{arxiv}
\usepackage{hyperref}
\usepackage[utf8]{inputenc} % allow utf-8 input
\usepackage[T1]{fontenc}    % use 8-bit T1 fonts
\usepackage{hyperref}       % hyperlinks
\usepackage{url}            % simple URL typesetting
\usepackage{booktabs}       % professional-quality tables
\usepackage{amsfonts}       % blackboard math symbols
\usepackage{nicefrac}       % compact symbols for 1/2, etc.
\usepackage{microtype}      % microtypography
\usepackage{graphicx}
\usepackage{tabularx}
\usepackage{multirow}
\usepackage{parskip}
\usepackage{float}
\usepackage{subcaption}
\usepackage{graphicx}
\usepackage{enumitem}
\title{Efficient Resource Management in Cloud Environment }
\date{}   
\author{ Smruti Rekha Swain$^a$,
Ashutosh Kumar Singh$^a$, Chung Nan Lee$^b$ \\
$^a$Department of Computer Applications, National Institute of Technology,\\
    Kurukshetra, Haryana-136119, India\\
$^b$Department of Computer Science and Engineering, National Sun Yat-sen University, \\ Kaohsiung, 804201, Taiwan
}

\hypersetup{
pdftitle={A template for the arxiv style},
pdfsubject={q-bio.NC, q-bio.QM},
pdfauthor={David S.~Hippocampus, Elias D.~Striatum},
pdfkeywords={First keyword, Second keyword, More},
}

\begin{document}

\maketitle
\renewcommand{\abstractname}{}
\textbf{Abstract.}

In cloud computing resource management plays a significant role in data centres and it is directly dependent on the application workload. Various services such as Infrastructure as a Service (IaaS), Platform as a Service (PaaS), and Software as a Service (SaaS) are offered by cloud computing to provide compute, network, and storage capabilities to the cloud users utilizing the pay-per-usage approach. Resource allocation is a prior solution to address various demanding situations like the under/overload handling, resource wastage, load balancing, Quality-of-Services (QoS) violations, VM migration and many more. The primary aim of Virtual Machine Placement (VMP) is mapping of Virtual Machines (VMs) to physical machines (PMs), such that the PMs may be utilized to their maximum efficiency, where the already active VMs are not to be interrupted. It provides a list of live VM migrations that must be accomplished to get the optimum solution and reduces energy consumption to a larger extent. The inefficient VMP leads to wastage of resources, excessive energy consumption and also increase overall operational cost of the data center. On this context, this article provides an extensive survey of resource management schemes in cloud environment. A conceptual scheme for resource management, grouping of current machine learning based resource allocation strategies, and fundamental problems of ineffective distribution of physical resources are analyzed. Thereafter, a complete survey of existing techniques in machine learning based mechanisms in the field of cloud resource management are explained. Ultimately, the paper explores and concludes distinct approaching challenges and future research guidelines associated to resource management in cloud environment.

\section{Introduction}
Cloud computing (CC) is an imminent distributed computing model that offers on-demand and low-cost shared computing resources to users \cite{ref1001}-\cite{ref1005}. It is an appropriate platform for large amount of data processing due to its on-demand elasticity or flexibility, incredibly low-latency and highly scalable processing capabilities\cite{ref1006}\cite{ref1007}. CC is an inevitable distributed computing model that offers on-demand and low-cost shared computing resources to users\cite{ref1008}\cite{ref1009}. It is an appropriate platform for a large amount of data processing due to its on-demand elasticity or flexibility, extremely low latency, and massively parallel processing features \cite{ref1010}. To deal with the ever growing and dynamic requests of cloud clients, it is necessary to achieve the Quality of Service (QoS) level among cloud clients and their cloud service providers (CSP) \cite{ref1011}. The CSPs must ensure proper synchronization among resource and power management within the data center while maintaining the required QoS for hosted applications. The efficiency of a data center substantially depends upon how VMs are sustained and where they are placed \cite{ref1012}\cite{ref1013}. The process of selecting the most appropriate physical machine (PM) in large cloud data centers is termed as VMP mechanism. The high power consumption inside the data center is due to the low utilization of computing resources such as CPU, memory, etc \cite{ref1014}-\cite{ref1016}. In the idle state, the server machines consume about 60\%-70\% power of their maximum power. High-energy consumption leads to an increase in the overall operational cost of the data center\cite{ref1017}-\cite{ref1020}. CC models are classified into two categories, i.e., the service model and the deployment model. Organizations choose a specific model based on their requirement.
Fig. 1 shows the conceptual resource management framework in a cloud environment.

%%%%%%%%%%%%%%%%%%%%%%%%%%%%%%%%%%%%%%%%%%%%%%%%%%%%%%%%%%%%%%%%%%%%%%%%%%%%%
 \begin{figure}[htp]
    \centering
    \includegraphics[width=0.95\textwidth]{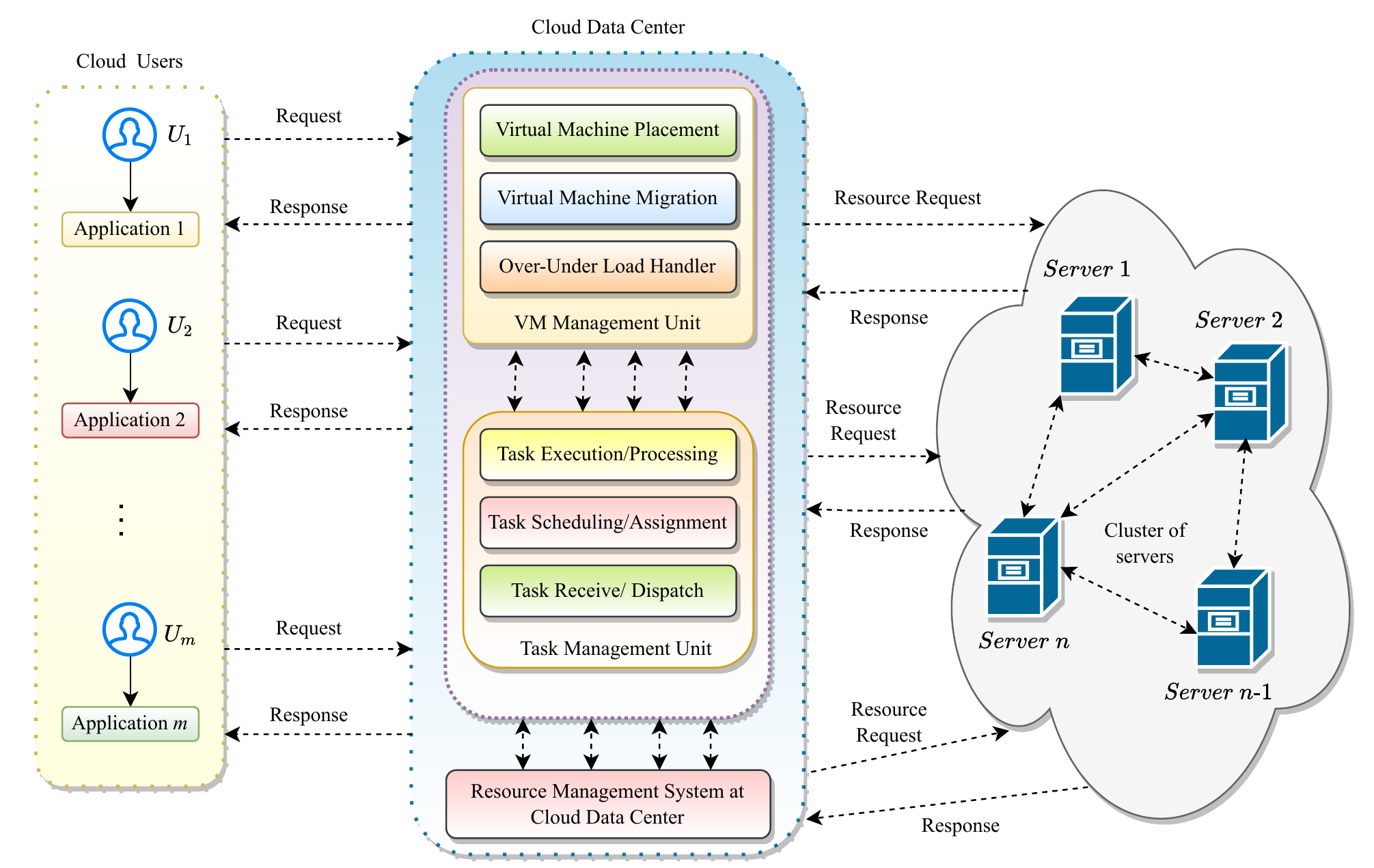}
    \caption{Conceptual Resource Management Framework in a Cloud Environment}
\end{figure}

%%%%%%%%%%%%%%%%%%%%%%%%%%%%%%%%%%%%%%%%%%%%%%%%%%%%%%%%%%%%%%%%%%%%%%%%%%%%%%%%%%%%%%%
The process of allocating various resources, i.e., memory, storage, and bandwidth, to a set of applications is termed resource management\cite{ref1021}. The Cloud Service Providers (CSPs) must ensure efficient and effective resource utilization by following the Service Level Agreements (SLAs) with the cloud users. Cloud data centers use virtualization technology in order to get minimum energy usage and maximum resource usage. Virtualization allows to sharing of the resources of the same PM among various VMs \cite{ref1022}. Depending upon the incoming demands list of VMs can be dynamically allocated and de-allocated in a single PM. Each VM has its own characteristics and consumes a distinct amount of power based on resource usage. Server consolidation approaches are used to consolidate the workloads on a fewer number of machines, and unused resources can be switched to a low-power mode or can be turned off to save energy \cite{ref1023}. Efficient VM mechanisms assign VMs to PMs in such a way as to increase resource usage, which reduces the number of active PMs. Hence, minimum energy consumption is achieved. Fig 2. shows the details of cloud computing models. 
%%%%%%%%%%%%%%%%%%%%%%%%%%%%%%%%%%%%%%%%%%%%%%%%%%%%%%%%%%%%%%%%%%%%%%%%%%%%%%%%%%
\begin{figure}[htp]
    \centering
    \includegraphics[width=0.99\textwidth]{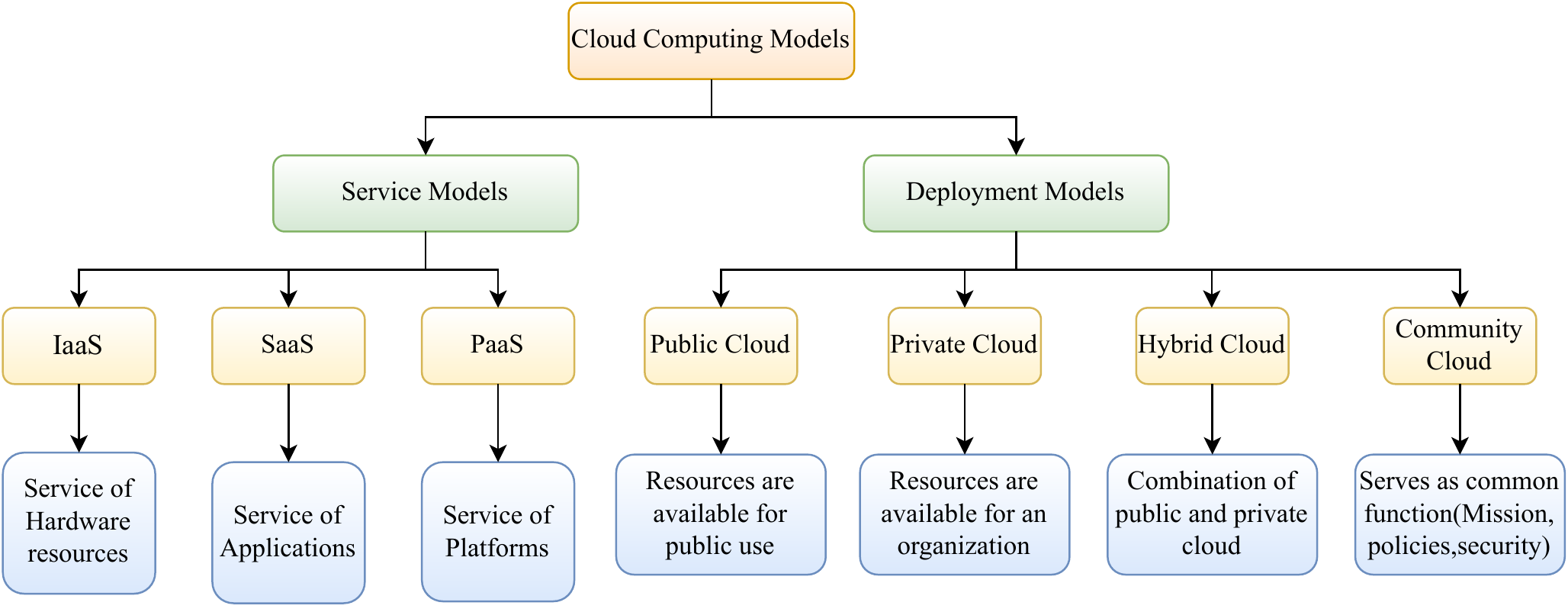}
    \caption{ Cloud Computing Models}
\end{figure}

%%%%%%%%%%%%%%%%%%%%%%%%%%%%%%%%%%%%%%%%%%%%%%%%%%%%%%%%%%%%%%%%%%%%%%%%%%%%%%%%%%%%%%%%%%%%%%%%%%%%%%
Resource management of cloud data centers is the major challenge which can be classified into three different levels, i.e., workload management at the application level, virtual machine level, and physical machine levels, as shown in Fig. 3. At the application level, during task scheduling and task assignment operations, various management techniques are required. Task assignment refers to the selection of appropriate VMs for task execution, and task scheduling deals with the selection of a task for its execution.
Task scheduling and task assignment can be achieved by using various mechanisms, as shown in Fig. 3. At the virtual machine level, workload management can be divided into three categories, i.e., VM assignment, VM consolidation, and VM assignment. The process of selecting the most appropriate PMs for VM is referred as VM assignment or VM placement. VM migration is referred as moving VMs from over/under-loaded PMs to a selected optimal PM, and VM consolidation is a technique to place VMs on the minimum number of PMs in order to reduce energy consumption and improve resource utilization. Various VM placement and VM migration approaches are shown in Fig. 3. Server consolidation,  over/under-load handling at the server, and resource provisioning are the primary operations associated with resource management at the physical level. The server consolidation is a technique that minimizes the number of active PMs by using power-efficient VMP approaches. The over/under-load handling deals with the identification of respective conditions at the server and mitigates its impact by using VM migration. The efficient distribution of physical resources among VMs before the arrival of the actual workload from cloud users is referred as resource provisioning. Different under/over-load handling mechanisms and resource provisioning approaches are illustrated in Fig. 3.
%%%%%%%%%%%%%%%%%%%%%%%%%%%%%%%%%%%%%%%%%%%%%%%%%%%%%%%%%%%%%%%%%%%%%%%%%%%%%%%%%%%%%%%%%%%%%%%%%%%%%%%%%
\begin{figure}[htp]
    \centering
    \includegraphics[width=0.95\textwidth]{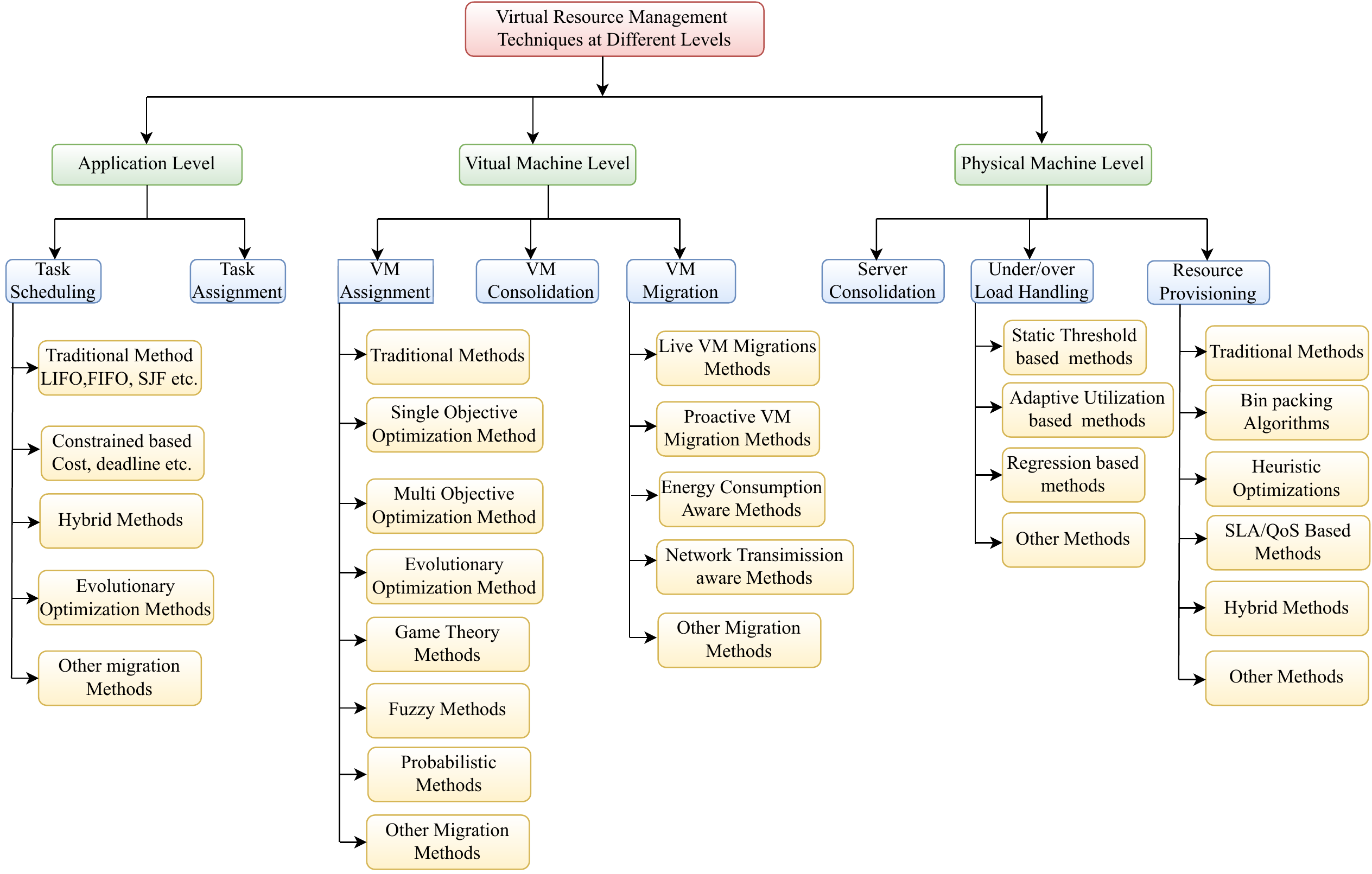}
    \caption{ Resource Management Techniques}
\end{figure}
%%%%%%%%%%%%%%%%%%%%%%%%%%%%%%%%%%%%%%%%%%%%%%%%%%%%%%%%%%%%%%%%%%%%%%
\section{Motivation}
To deal with the ever-growing and dynamic demands of cloud clients, a huge number of VMs are placed on more PMs. The VMP performs a crucial role in power consumption, resource utilization, and the overall operational cost of the data center.
High power consumption within the data center is due to inefficient use of these resources. Excess energy consumption by cloud data centers facilities high operational costs and, in turn, increases carbon dioxide ($CO_2$) levels indirectly. It has been noted that cooling systems and PMs collectively use approximately 70\% of the overall power used in a data center \cite{ref101}\cite{ref1018}. Power consumption, resource utilization and carbon foot print are fundamental components for CSPs to control operational costs. Due to unbalanced VM placement, most of the time, it has been observed that resources are either underutilized or overloaded, which leads to wastage of resources and compromises the quality of service parameters. The server machines consume about 60\%-70\% power of their maximum power in the idle state \cite{ref102}\cite{ref1020}. Server consolidation mechanisms are used to consolidate the workloads on a lesser number of machines, and unused resources can be switched to a low-power mode, e.g., using DVFS \cite{ref103}\cite{ref1022} or can be turned off to save energy. Various energy-saving approaches reduce the performance of the system and create a complex balance between power savings and high performance. It shows the overload and under-utilization of resources against the typical resource capacity of a data center. Some of the key challenges due to inefficient distribution of physical resources at the data center are excess power consumption, resource wastage, performance degradation, frequent VM migration, increased SLA violations, and minimized QoS. To overcome these issues, the data centers must have ideal resource constraint capacity in order to satisfy the incoming loads without wasting resources. Again, the resource manager can achieve this if the accurate forecasting of future demand is known beforehand. This emphasizes the requirement for automatic resource management strategies that allows systems to auto-adapt based on variable resource demands by efficiently using the current resources.

\section{Related Work}
 Multi-objective energy efficient VMP\\
 Sharma et al. \cite{ref8} proposed an energy-efficient multi-objective VMP framework to minimize power consumption, maximize resource utilization and reduce SLA violation. The integration of genetic algorithm (GA) and particle swarm optimization (PSO) termed the HGAPSO algorithm was used to assign VMs on PMs. The migration of VMs from inefficient to the optimal PM is accomplished by GA however, it has poor convergence. So, PSO assists GA in choosing the most optimal PM within data center. The drawback of this approach is it is not applicable in a heterogeneous environment.\\

Secure and Energy Aware Load Balancing (SEA-LB)\\
 Singh et al. \cite{ref9} suggested a secure and energy aware load balancing (SEA-LB) scheme for VMP problem. This approach use GA to improve resource utilization, minimize power consumption and decreasing number of conflicting server by providing security to each VM within the data center during VMP. The lack of this scheme is that it generally leads to untimely convergence.\\
 
Secure and Multi-objective VM Placement\\
Saxena et al. \cite{ref4} introduced a Secure and Multi-objective Virtual Machine Placement (SM-VMP) mechanism based on NSGA-II and whale optimization evolutionary technique. This work focused on reduction in security threats, energy consumption, communication cost and enhanced resource utilization simultaneously. A probability based encoding and decoding scheme was suggested to encode VM allocations as a whale position vector. To check the efficacy of this approach, the obtained outcomes are compared with various existing multi-objective VMPs such as HGAPSO, Non-dominated based GA (NSGA-II), Random-fit, First-fit and Best-fit.\\

Multi-objective Load Balancing\\
Saxena et al.\cite{ref7} introduced a multi-objective load-balancing (OP-MLB) technique. It is a power-efficient and online resource prediction based approach. Evolutionary neural-network used to predict future demands of VMs. Proactive under/over-load identification on PMs is carried out and NSGA-II algorithm is used for VM placement. An energy and network efficient VM migration technique was used to reduce power consumption within the data center. It permits energy saving by switching off the idle PMs and decreasing the number of active PMs, minimizing VM migrations and improving the resource utilization.\\

Multi-input and Multi-output Neural Network\\
Saxena et al. \cite{ref6} proposed a multi-resource feed-forward evolutionary neural network based prediction technique by applying modifications of an existing single input and single output (SISO) feed-forward neural network. To collect multiple inputs and predict output based on multiple attributes there are sets of neural nodes in stead of conventional nodes at input, hidden and output layers. This approach enhanced prediction accuracy with nominal requirement of time complexity and space complexity as compared to SISO neural networks. This incorporated technique optimized resource utilization and energy consumption of the data center.\\

Quantum Neural Network\\
Singh et al. \cite{ref10} suggested a resource prediction model for cloud environment. This work uses Evolutionary Quantum Neural Network (EQNN) which is the combination of precise concepts of quantum computing and evolutionary algorithm. Quantum gates such as rotation gates and C-NOT gates are used as an activation function inside the neural network. During training phase, there is a assumption that qubits have superior exploration and exploitation capacity than actual numbers. Qubits are generated from pre-processed training input data with the help of rotation gate. Then, qubit values are generated for network connections instead of real-numbered weights.\\ 

Evolutionary Neural Network\\
A workload prediction model suggested by Kumar et al.\cite{ref3} for cloud environment by applying neural network and self adaptive differential evolution (SaDE) algorithm. This approach deals with collection of historical data from various cloud users then  pre-processing over these data is conducted by normalizing it in the range of [0, 1]. Here, prediction model predict future workload on data center at n + 1 time instance by extracting patterns from actual workload and analysing of n previous workload values. Then, SaDE algorithm is applied to train the neural network. The more accurate results are obtained due to fast convergence, higher exploration, and learning capability of the SaDE algorithm.\\

Kumar et al. \cite{ref2} proposed a workload prediction approach that is based on neural network model with supervised learning mechanism. A biphase adaptive differential evolution (BaDE) learning algorithm used to train the neural network which motivated by dual adaptation such as, at crossover level during exploitation process and mutation in exploration phase to enhance the performance of neural network. This technique provided better prediction accuracy.\\

Long Short-term Recurrent Neural Network (LSTM-RNN)\\
Long short-term memory model in a recurrent neural network (LSTM-RNN) proposed by Kumar et al.\cite{ref1} for load prediction of fine-grained host. This model use the Back-propagation algorithm between recurrent layers and obtains better prediction accuracy for server loads. The major drawback of this approach is that, during the training phase it takes long computational time.\\

Communication Cost Aware Resource Efficient Load Balancing\\
Saxena et al. \cite{ref5}  proposed a communication cost aware and resource efficient load balancing (CARELB) scheme to minimize communication cost among VMs, reduce power consumption and improve resource utilization of the data center. VMs
are assigned with high affinity and inter-dependency physically nearer to each other.  In order to obtain optimal VM allocations a combined concept of Particle Swarm Optimization and NSGA-II i.e. PSOGA algorithm was used.\\

Complex-Valued Neural Network\\
 Qazi et al. \cite{ref11} proposed a multi-layered neural networks with multi-valued neurons (MLMVN) prediction model. A complex-valued neural network with a derivative-free feed-forward learning  was used by this approach. Since, complex-valued neural network is more efficient than traditional real-valued based neural network. It allows high learning capability and better accuracy in minimum time.\\

Multi objective VM placement\\
A workload prediction and allocation technique suggested in \cite{ref12}. It is a multi-objective allocation approach that deal with memory and CPU utilization of VMs and PMs ensures minimum power usage within the data center. The GA was used to predict future resource usage before assigning VMs to PMs. After that VMP algorithm was employed to enhance resource utilization and minimize power consumption.

%%%%%%%%%%%%%%%%%%%%%%%%%%%%%%%%%%%%%%%%%%%%%%%%%%%%%%%%%%%%%%
\begin{table} [!htbp]
    \centering
    \caption{ Comparative Summary} 
   % \resizebox{\columnwidth}{!}{
    \begin{tabular}{ |p{3cm}|p{3cm}|p{3cm}|p{2cm}|} \hline
    Author, Objective & Approach & Strength & Weakness\\  \hline 
    Sharma et al., 2016 \cite{ref8}, Multi-objective VM placement & Hybrid of PSO and GA algorithm & Maximizes resource utilization and minimizes power consumption & Resource wastage\\ \hline
    Singh et al., 2019 \cite{ref9}, Multi-objective VM placement &  NSGA-II & Consideration of security during load balancing & VM migration is not considered \\ \hline
    Saxena et al., 2021 \cite{ref4}, Multi-objective VM placement & Whale Optimization Algorithm and NSGA-II  & Maximize resource utilization, minimize energy consumption, communication cost and security &  Cluster level VM mapping is not considered \\ \hline
    Saxena et al., 2021 \cite{ref7}, online predictive resource management & multi-objective VM placement based on Online VM resource-usage using evolutionary optimization mechanisms & Energy-efficient VM placement with proactive handling of overload & inaccurate overload prediction may lead to unnecessary VM migration\\ \hline
   Saxena et al., 2021 \cite{ref6},  VM auto scaling &  Workload prediction and clustering  technique is used to get accurate number and size of VMs  & Energy efficient VM placement &  under-loaded servers are ignored\\ \hline
   Singh et al. ,2021 \cite{ref10}, workload prediction model & Quantum based evolutionary
   neural network optimized by Self-balanced adaptive differential evolutionary algorithm & Higher and stable prediction
   accuracy & Higher complexity because of generation and processing of qubits \\ \hline
   Kumar et al., 2018 \cite{ref3}, workload prediction model & Artificial Neural Network
    trained with SaDE evolutionary algorithm & enhanced prediction accuracy over Back-propagation algorithm & Static approach for prediction \\ \hline
    Kumar et al., 2018 \cite{ref1}, workload prediction model & Long short-term memory model in a recurrent neural network (LSTM-RNN) & better accuracy than ANN & Long Computational time  \\ \hline
    kumar et al., 2020  \cite{ref2} &  Bi-phase adaptive differential evolution algorithm used to optimize neural network  & Faster Convergence & Higher complexity than SaDE algorithm optimized neural network  \\ \hline
    Saxena et al., 2020 \cite{ref5}, Communication-cost aware VM placement & Integration of PSO and NSGA-II
    algorithm (PSOGA) & enhances resource utilization and minimizes response-time &  VM mapping at cluster-level is not considered \\ \hline
    
    \end{tabular}
    
    \label {table:1}
    
    \end{table}

%%%%%%%%%%%%%%%%%%%%%%%%%%%%%%%%%%%%%%%%%%%%%%%%%%%%%%
\section{Emerging Challenges and Future Research Directions}

In the light of current resource management mechanisms, the subsequent issues are summarized for the improvement and advancement of more promising resource management framework in cloud environment:
\begin{enumerate}
    \item \textit{Enhancing accuracy of Resource prediction Model :}  In order to avoid SLA violations and permit contention free execution of user task, it is necessary to improve the VM resource prediction accuracy. The accurate prediction of future resource requirement provides reliable and highly available services to cloud users. Most of the existing techniques are static and deals with ancient workload only.
    \item \textit{Overloading and overall Performance Degradation:} In overload condition, the collective resource demand of VMs exceeds the available resource capacity of the PMs leading to overloaded PMs. It results overall performance degradation and SLA violations such as, unavailability of resources, longer response time, some VMs may crash etc.
    \item \textit{Live migration of VMs and Downtime:} To handle the variable user demands and arbitrary requirement of excess resource capacities or to handle overload situations, VMs are switched from overloaded PMs to most optimal PM which has ample resource capacity that leads to delayed execution. Another primary challenge is to decide which VM should migrate, when to migrate, and to which PM it should migrate?
    \item \textit{Lack of security during resource allocation:} Security is the major concern of every cloud user. Resource management policies must ensure the distribution of physical resources among various users’ VMs must consider the security of data as a principal perspective along with capacity constraint of resources.
    \item \textit{Consideration of single resource during load balancing:} Most of the current elastic resource distribution strategies deals with single resource such as CPU only, whereas all the resources are significant as contention of either of the resources i.e., memory, bandwidth, disk usage, processor may become bottleneck degrading overall performance of the system in real world cloud resource management.
    \item \textit{Interaction and Co-operation among associated operations:} Separate consideration ofresource prediction, resource provisioning, VM placement and VM migration, nevertheless all of these operations provides common objective of energy-efficient resource utilization. Therefore, all the related resource management operations must be employed at homogeneous platform to permit interaction among them for advancement of a realistic resource management solution in cloud environment.
\end{enumerate}

The subsequent research guidelines are designed which will fill the research gaps and deal with aforesaid challenges:
\begin{enumerate}
    \item  An enhanced resource prediction technique with maximum accuracy is required to be developed. The resource predictor should analyze dynamically in real-time by training and re-training itself with ancient data along with real-time resource usage  by distinct VMs and PMs. Additionally, multiple resources must be expected simultaneously by analyzing utilization of various resources from the training data samples.
    \item  An efficient over/under-load forecasting technique is necessary which can primitively compute the high and down load situation with maximum accuracy that can allow migration of VMs before occurrence of the under/over-load and reduce performance degradation.
    \item To optimize multiple resources usage such as memory, CPU, bandwidth simultaneously throughout load balancing  for contention-free execution of users’ task while preserving power, decreasing resource wastage, minimizing carbon emission in cloud data center.
    \item To develop a secure resource distribution schemes for cloud environment which permits secure execution of users task, secure inter communication between interdependent VMs.
    \item To develop a reliable fault tolerant framework to enhance the resource utilization and meet QoS parameter.
    \item  To incorporate workload prediction technique with an efficient load balancing approach at systematic platform to improve the effectiveness during resource management.

\end{enumerate}


\begin{thebibliography}{20}

\bibitem{ref1001}
Kumar, Jitendra, and Ashutosh Kumar Singh. "Dynamic resource scaling in cloud using neural network and black hole algorithm." In 2016 Fifth International Conference on Eco-friendly Computing and Communication Systems (ICECCS), pp. 63-67. IEEE, 2016.

\bibitem{ref1002}
Kumar, Jitendra, and Ashutosh Kumar Singh. "Cloud datacenter workload estimation using error preventive time series forecasting models." Cluster Computing 23, no. 2 (2020): 1363-1379.

\bibitem{ref1003}
Gupta, Ishu, Rishabh Gupta, Ashutosh Kumar Singh, and Rajkumar Buyya. "MLPAM: A machine learning and probabilistic analysis based model for preserving security and privacy in cloud environment." IEEE Systems Journal 15, no. 3 (2020): 4248-4259.
    
\bibitem{ref1004}
Gupta, Rishabh, and Ashutosh Kumar Singh. "Differential and Access Policy Based Privacy-Preserving Model in Cloud Environment." Journal of Web Engineering (2022): 609-632.
    
\bibitem{ref1005}
Gupta, R. and Singh, A.K., 2022, March. Privacy-Preserving Cloud Data Model based on Differential Approach. In 2022 Second International Conference on Power, Control and Computing Technologies (ICPC2T) (pp. 1-6). IEEE.

\bibitem{ref1006}
Chhabra, Sakshi, and Ashutosh Kumar Singh. "Dynamic Resource Allocation Method for Load Balance Scheduling Over Cloud Data Center Networks." Journal of Web Engineering (2021): 2269-2284.
    
\bibitem{ref1007}
Chhabra, Sakshi, and Ashutosh Kumar Singh. "A secure VM allocation scheme to preserve against co-resident threat." International Journal of Web Engineering and Technology 15, no. 1 (2020): 96-115

\bibitem{ref1008}
Kumar, Jitendra, Ashutosh Kumar Singh, and Anand Mohan. "Resource‐efficient load‐balancing framework for cloud data center networks." ETRI Journal 43, no. 1 (2021): 53-63.

\bibitem{ref1009}
Gupta, Rishabh, Deepika Saxena, and Ashutosh Kumar Singh. "Data security and privacy in cloud computing: concepts and emerging trends." arXiv preprint arXiv:2108.09508 (2021).

\bibitem{ref1010}
Chhabra, S., and A. K. Singh. "Dynamic hierarchical load balancing model for cloud data centre networks." Electronics Letters 55, no. 2 (2019): 94-96.

\bibitem{ref1011}
Chhabra, Sakshi, and Ashutosh Kumar Singh. "OPH-LB: Optimal Physical Host for Load Balancing in Cloud Environment." Pertanika Journal of Science \& Technology 26, no. 3 (2018).

\bibitem{ref1012}
Singh, Ashutosh Kumar, and Rishabh Gupta. "A privacy-preserving model based on differential approach for sensitive data in cloud environment." Multimedia Tools and Applications (2022): 1-24.

\bibitem{ref1013}
Chhabra, Sakshi, and Ashutosh Kumar Singh. "Optimal VM placement model for load balancing in cloud data centers." In 2019 7th International Conference on Smart Computing \& Communications (ICSCC), pp. 1-5. IEEE, 2019.

\bibitem{ref1014}
Kumar, Jitendra, and Ashutosh Kumar Singh. "Cloud resource demand prediction using differential evolution based learning." In 2019 7th International Conference on Smart Computing \& Communications (ICSCC), pp. 1-5. IEEE, 2019.

\bibitem{ref1015}
Kumar, Jitendra, and Ashutosh Kumar Singh. "Performance assessment of time series forecasting models for cloud datacenter networks’ workload prediction." Wireless Personal Communications 116, no. 3 (2021): 1949-1969.

\bibitem{ref1016}
Saxena, Deepika, and Ashutosh Kumar Singh. "VM Failure Prediction based Intelligent Resource Management Model for Cloud Environments." In 2022 Second International Conference on Power, Control and Computing Technologies (ICPC2T), pp. 1-6. IEEE, 2022.

\bibitem{ref1017}
Kumar, Jitendra, and Ashutosh Kumar Singh. "Adaptive Learning based Prediction Framework for Cloud Datacenter Networks' Workload Anticipation." Journal of Information Science \& Engineering 36, no. 5 (2020).


\bibitem{ref1018}
Saxena, Deepika, and A. K. Singh. "Security embedded dynamic resource allocation model for cloud data centre." Electronics Letters 56, no. 20 (2020): 1062-1065.

\bibitem{ref1019}
Saxena, Deepika, and Ashutosh Kumar Singh. "Auto-adaptive learning-based workload forecasting in dynamic cloud environment." International Journal of Computers and Applications (2020): 1-11.

\bibitem{ref1020}
 Saxena, Deepika, and Ashutosh Kumar Singh. "OSC-MC: Online secure communication model for cloud environment." IEEE Communications Letters 25, no. 9 (2021): 2844-2848.



\bibitem{ref1021}
Saxena, Deepika, and Ashutosh Kumar Singh. "Energy aware resource efficient-(eare) server consolidation framework for cloud datacenter." In Advances in communication and computational technology, pp. 1455-1464. Springer, Singapore, 2021.

\bibitem{ref1022}
Saxena, Deepika, and Ashutosh Kumar Singh. "OFP-TM: an online VM failure prediction and tolerance model towards high availability of cloud computing environments." The Journal of Supercomputing 78, no. 6 (2022): 8003-8024.

\bibitem{ref1023}
Saxena, Deepika, Ishu Gupta, Ashutosh Kumar Singh, and Chung-Nan Lee. "A Fault Tolerant Elastic Resource Management Framework Towards High Availability of Cloud Services." IEEE Transactions on Network and Service Management (2022).

\bibitem{ref101}
Eugen Feller, Louis Rilling, and Christine Morin. Energy-aware ant colony based workload placement in clouds. In
2011 IEEE/ACM 12th International Conference on Grid Computing, pages 26–33. IEEE, 2011.

\bibitem{ref102}
Neha Garg, Damanpreet Singh, and Major Singh Goraya. Power and resource-aware vm placement in cloud
environment. In 2018 IEEE 8th International Advance Computing Conference (IACC), pages 113–118. IEEE, 2018.

\bibitem{ref103}
Li Shang, Li-Shiuan Peh, and Niraj K Jha. Dynamic voltage scaling with links for power optimization of interconnec-
tion networks. In The Ninth International Symposium on High-Performance Computer Architecture, 2003. HPCA-9 2003. Proceedings., pages 91–102. IEEE, 2003.



\bibitem{ref1}
Jitendra Kumar, Rimsha Goomer, and Ashutosh Kumar Singh. Long short term memory recurrent neural network (lstm-rnn) based workload forecasting model for cloud datacenters. Procedia Computer Science, 125:676–682, 2018.

\bibitem{ref2}
Jitendra Kumar, Deepika Saxena, Ashutosh Kumar Singh, and Anand Mohan. Biphase adaptive learning-based neural network model for cloud datacenter workload forecasting. Soft Computing, 24(19):14593–14610, 2020.

\bibitem{ref3}
Jitendra Kumar and Ashutosh Kumar Singh. Workload prediction in cloud using artificial neural network and adaptive differential evolution. Future Generation Computer Systems, 81:41–52, 2018.

\bibitem{ref4}
Deepika Saxena, Ishu Gupta, Jitendra Kumar, Ashutosh Kumar Singh, and Xiaoqing Wen. A secure and multiobjective virtual machine placement framework for cloud data center. IEEE Systems Journal, 2021.

\bibitem{ref5}
Deepika Saxena and Ashutosh Kumar Singh. Communication cost aware resource efficient load balancing (care-lb)
framework for cloud datacenter. Recent Advances in Computer Science and Communications, 12:1–00, 2020.

\bibitem{ref6}
Deepika Saxena and Ashutosh Kumar Singh. A proactive autoscaling and energy-efficient vm allocation framework
using online multi-resource neural network for cloud data center. Neurocomputing, 426:248–264, 2021.

\bibitem{ref7}
Deepika Saxena, Ashutosh Kumar Singh, and Rajkumar Buyya. Op-mlb: An online vm prediction based multi-
objective load balancing framework for resource management at cloud datacenter. IEEE Transactions on Cloud
Computing, 2021.

\bibitem{ref8}
Neeraj Kumar Sharma and G Ram Mohana Reddy. Multi-objective energy efficient virtual machines allocation at
the cloud data center. IEEE Transactions on Services Computing, 12(1):158–171, 2016.

\bibitem{ref9}
Ashutosh Kumar Singh and Jitendra Kumar. Secure and energy aware load balancing framework for cloud data
centre networks. Electronics Letters, 55(9):540–541, 2019.

\bibitem{ref10}
Ashutosh Kumar Singh, Deepika Saxena, Jitendra Kumar, and Vrinda Gupta. A quantum approach towards the adaptive prediction of cloud workloads. IEEE Transactions on Parallel and Distributed Systems, 32(12):2893–
2905, 2021.

\bibitem{ref11}
Kashifuddin Qazi and Igor Aizenberg. Cloud datacenter workload prediction using complex-valued neural networks.
In 2018 IEEE Second International Conference on Data Stream Mining \& Processing (DSMP), pages 315–321. IEEE,
2018.

\bibitem{ref12}
Fan-Hsun Tseng, Xiaofei Wang, Li-Der Chou, Han-Chieh Chao, and Victor CM Leung. Dynamic resource prediction
and allocation for cloud data center using the multiobjective genetic algorithm. IEEE Systems Journal, 12(2):1688–
1699, 2017.

\end{thebibliography}
\end{document}